# Exploring Cation Selection and Disorder within Entropy-Driven $A_6B_2O_{17}$ (*A*=Zr/Hf, *B*=Nb/Ta) Oxides

*Jacob T. Sivak[1], R. Jackson Spurling[2], Jon-Paul Maria[2], *Susan B. Sinnott[1,2,3]

[1]*Department of Chemistry, The Pennsylvania State University, University Park, PA 16802, USA*

[2]*Department of Materials Science and Engineering, The Pennsylvania State University, University Park, PA 16802, USA*

[3]*Institute for Computational and Data Sciences, The Pennsylvania State University, University Park, PA 16802, USA*

**Corresponding Authors:** Jacob T. Sivak jts6114@psu.edu and Susan B. Sinnott sbs5563@psu.edu

## Abstract

We investigate the local atomic and electronic structure, thermodynamic stability, and defect chemistry of $A_6B_2O_{17}$ (*A* = Zr/Hf, *B* = Nb/Ta) oxides using first-principles density functional theory (DFT) calculations. We examine both ordered unit cells as well as fully disordered special quasirandom structures to clearly discern the effects of cation disorder. Structural predictions align closely with previous experimental results and follow established ionic radii trends. The electronic structure is strongly dependent on *B*-cation species: $A_6Ta_2O_{17}$ compositions have ~30% larger band gaps than their $A_6Nb_2O_{17}$ counterparts. Defect chemistry is similar for all compositions, with anion vacancies being more energetically favorable than corresponding cation defects. All explored $A_6B_2O_{17}$ compositions are enthalpically unstable with respect to their $AO_2$ and $B_2O_5$ competing oxides and are therefore classified as entropy-stabilized materials, supporting prior experimental results. The pronounced agreement between our disordered supercell predictions with experimental measurements indicates all explored $A_6B_2O_{17}$ compositions contain substantial cation disorder across all 6-, 7-, and 8-coordinated sites. Our findings collectively provide a fundamental understanding of the $A_6B_2O_{17}$ material family through DFT calculations, establishing a framework for future compositional tuning to engineer targeted material properties.

# Introduction

Entropy-driven materials discovery enables atypical local chemical environments with unique structure-property relationships unavailable through traditional, enthalpy-driven approaches. The work of Rost et al. in 2015 demonstrates this stabilization paradigm in which the high-symmetry rock salt lattice is randomly decorated with five cations to form the single-phase $Mg_{1/5}Co_{1/5}Ni_{1/5}Cu_{1/5}Zn_{1/5}O$ composition [1]. This pioneering work has since inspired the discovery of many other high-entropy oxides within various other crystal structures, with promising reports of advancing functional properties in energy storage [2,3], magnetism [4,5], optical absorption [6,7], and ionic conductivity [8]. While these materials commonly contain five or more components following the prototypical $Mg_{1/5}Co_{1/5}Ni_{1/5}Cu_{1/5}Zn_{1/5}O$ chemical recipe, such a large number of components is not explicitly needed to entropy-stabilize a disordered structure with unique local coordinations. Rather, entropy-stabilization solely requires that a material possess a positive enthalpy ($\Delta H$) barrier that is overcome by entropy ($\Delta S$) to result in a negative Gibbs free energy ($\Delta G = \Delta H - T\Delta S$) above some critical temperature ($T$) [9,10]. Entropy-stabilization is therefore possible in a chemically disordered oxide containing as little as two cations such as the copper-containing spinels noted by Navrotsky and Kleppa roughly forty years prior to Rost et al.'s $Mg_{1/5}Co_{1/5}Ni_{1/5}Cu_{1/5}Zn_{1/5}O$ discovery [11].

The $A_6B_2O_{17}$ structure presents a particularly interesting case to test this lower-component entropy-stabilization hypothesis as its superlattice structure contains multiple unique cation environments (Figure 1) [12]. Assuming a fully random cation decoration across all cation polyhedra in the $A_6B_2O_{17}$ structure, the ideal configurational entropy per formula unit is 4.50R J/mol·K, roughly 2.8 times larger than that of the prototypical $Mg_{1/5}Co_{1/5}Ni_{1/5}Cu_{1/5}Zn_{1/5}O$ high-entropy oxide (1.61R) despite $A_6B_2O_{17}$ containing only two cations. While an ordered cation decoration was initially proposed by Galy and Roth [13], more recent experimental reports suggest substantial cation disorder for this structural family with random cation decoration across all 6-, 7-, and 8-coordinated lattice sites [12,14]. Calorimetry measurements for $Zr_6Nb_2O_{17}$, $Zr_6Ta_2O_{17}$, $Hf_6Nb_2O_{17}$, and $Hf_6Ta_2O_{17}$ additionally indicate positive enthalpy costs with respect to their binary oxide components [15], leading to $A_6B_2O_{17}$ ($A$ = Zr/Hf and $B$ = Nb/Ta) compositions being classified as entropy-stabilized oxides containing only two cation species.

(a) (b)

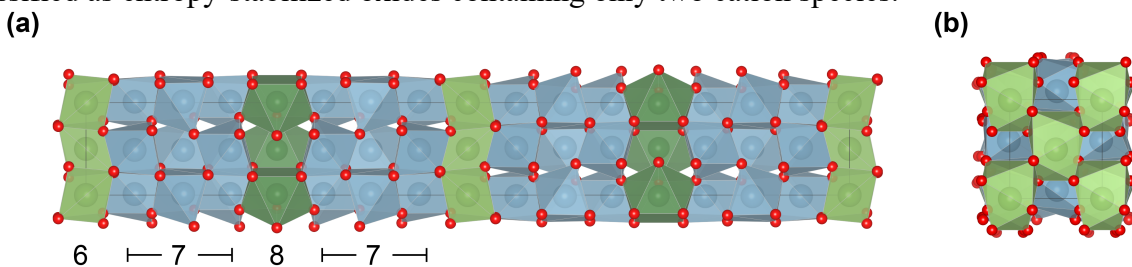

6 ⊢ 7 ⊣ 8 ⊢ 7 ⊣

Figure 1. $A_6B_2O_{17}$ unit cell looking down (a) [001] and (b) [010] directions. Polyhedra are colored and labeled based on the cation-oxygen coordination number. Structure obtained from McCormack et al. [12] and visualized used VESTA [16].

While significant experimental focus has been given to understanding the $A_6B_2O_{17}$ atomic structure, chemical disorder, and thermodynamic stability, there are also functional property opportunities for these formulations. High-temperature applications such as thermal barrier coatings have been the most extensively investigated as $Hf_6Ta_2O_{17}$ exhibits outstanding thermal and mechanical properties that rival the industry standard yttria-stabilized zirconia [17–20]. $A_6B_2O_{17}$ phases have also more recently been investigated as electronic materials, specifically dielectrics [21–23], highlighting additional opportunities for continued advancement. Much of the underlying phenomena associated with these properties remain unknown, however, as few density functional theory (DFT) studies have been performed due to the significant computational cost associated with the large 100-atom $A_6B_2O_{17}$ unit cell.

In this manuscript, we provide a self-consistent first-principles exploration to understand how cation selection influences the $A_6B_2O_{17}$ ($A$ = Zr/Hf and $B$ = Nb/Ta) local atomic and electronic structure, thermodynamic stability, and defect chemistry. We utilize the regularized-restored strongly constrained and appropriately normed (r$^2$SCAN) meta-generalized gradient approximation (meta-GGA) functional in our calculations for its improved predictive accuracy over the more commonly used local density approximation (LDA) and generalized gradient approximation (GGA) [24–26]. Additionally, we elucidate the effect that chemical disorder plays on resulting properties by comparing both the ordered and disordered cation decorated structures. Where possible, we compare our predictions to experimental measurements from literature, instilling confidence that DFT predictions can quantitatively capture chemical, structural, and stability differences among $Zr_6Nb_2O_{17}$, $Zr_6Ta_2O_{17}$, $Hf_6Nb_2O_{17}$, and $Hf_6Ta_2O_{17}$ compositions. Our findings collectively establish a foundational understanding of the $A_6B_2O_{17}$ entropy-driven materials family using DFT calculations to enable future compositional tuning for specific properties of interest.

## Ordered Cation Predictions

We begin exploring the $A_6B_2O_{17}$ materials family using the 100-atom unit cell with ordered cation decoration in which $A$-cations occupy 7-coordinated sites and $B$-cations occupy 6- and 8-coordinated sites (as shown in Figure 1) following the original structure solution proposed by Galy and Roth [13]. While experimental characterization has recently indicated that $A_6B_2O_{17}$ materials exhibit a disordered cation sublattice [12], beginning with the ordered unit cell enables us to establish the ability of our calculations to accurately model these systems as well as to disentangle the effect of chemical disorder by comparing ordered and disordered predictions. Unit cell volumes for all ordered $A$ = Zr/Hf and $B$ = Nb/Ta combinations ($Zr_6Nb_2O_{17}$, $Zr_6Ta_2O_{17}$, $Hf_6Nb_2O_{17}$, $Hf_6Ta_2O_{17}$) are presented in Figure 2. All compositions maintain the initial *Ima2* orthorhombic symmetry identified from experiment throughout the relaxation process [12]. Our $Hf_6Ta_2O_{17}$ volumes predicted with r$^2$SCAN differ from experimental measurements by only 0.35% [12], compared to the previous DFT calculations in the literature using LDA (3.3%) or GGA (3.6%) [27]. Following expectations from stoichiometry and tabulated ionic radii [28], $Zr_6B_2O_{17}$ compositions are ~2.9% larger than $Hf_6B_2O_{17}$, while $A_6Nb_2O_{17}$ are only ~0.4% larger than $A_6Ta_2O_{17}$ compositions. Our predictions align with d-spacing computed from X-ray diffraction (XRD) measurements for the relative 811 peak positions for this composition set [14]. Relaxed bond lengths across all three cation coordinations are shown in Figure S1. Bond length trends align with those found for the unit cell volumes: $Zr_6B_2O_{17}$ and $A_6Nb_2O_{17}$ compositions have longer bond lengths than their $Hf_6B_2O_{17}$ and $A_6Ta_2O_{17}$ counterparts, respectively. Importantly, all average bond lengths exist within a narrow window (<10%) for these ordered unit cells, aligning with known rules for solid-solution formation of chemically disordered materials [29,30].

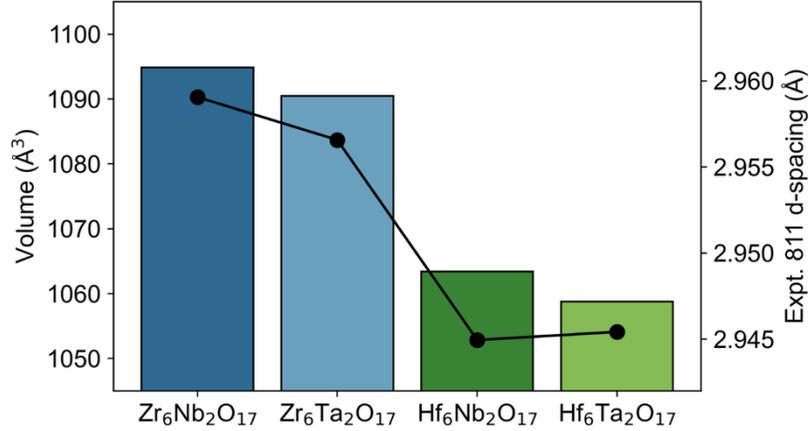

Figure 2. Calculated unit cell volume predictions for ordered $A_6B_2O_{17}$ ($A$ = Zr/Hf, $B$ = Nb/Ta) unit cells. Experimental 811 d-spacing values from Spurling et al. [14] are shown as black circles along the y-axis on the right side.

To explore how cation selection impacts the $A_6B_2O_{17}$ electronic structure, we calculate the projected band structures for all ordered unit cells (Figure 3). Electronic band gaps are all indirect and range from 2.36 – 3.25 eV, indicating insulating behavior for all compositions. All band structures are qualitatively similar: oxygen p-orbitals dominate the top of the valence band, while unoccupied $B$-cation d-orbitals occupy the bottom of the conduction band. $A$-cation d-orbitals reside higher in the conduction band, resulting in $A$-cation species having little effect on the band gap (< 0.1 eV), with $Hf_6B_2O_{17}$ compositions having slightly larger band gaps than their $Zr_6B_2O_{17}$ counterparts. $B$-cation species, however, have a significant influence on the $A_6B_2O_{17}$ electronic structure, resulting in a roughly 30% band gap increase for $A_6Ta_2O_{17}$ compositions (~3.2 eV) compared to $A_6Nb_2O_{17}$ compositions (~2.4 eV). We attribute the larger electronic band gaps of $A_6Ta_2O_{17}$ compositions to the higher ionic bonding character of Ta compared to Nb, which can be observed using the average Bader charge computed for each ion (Table S1). $A_6B_2O_{17}$ electronic structures generally maintain similar trends to their respective competing phases ($ZrO_2$, $HfO_2$, $Nb_2O_5$, $Ta_2O_5$), however exhibit slightly reduced band gaps (Figure S2). A larger band gap for $Hf_6Ta_2O_{17}$ (3.25 eV) is observed in our $r^2$SCAN calculations compared to the previously reported PBE band gap of 2.67 eV [27], supporting the higher predictive accuracy provided by meta-GGA functionals like $r^2$SCAN [24,25]. Trends within our first-principles predictions support recent experimental reports that $A_6Nb_2O_{17}$ compositions are more conductive than $A_6Ta_2O_{17}$, with $A$-cation species having little effect on conductivity [23,31].

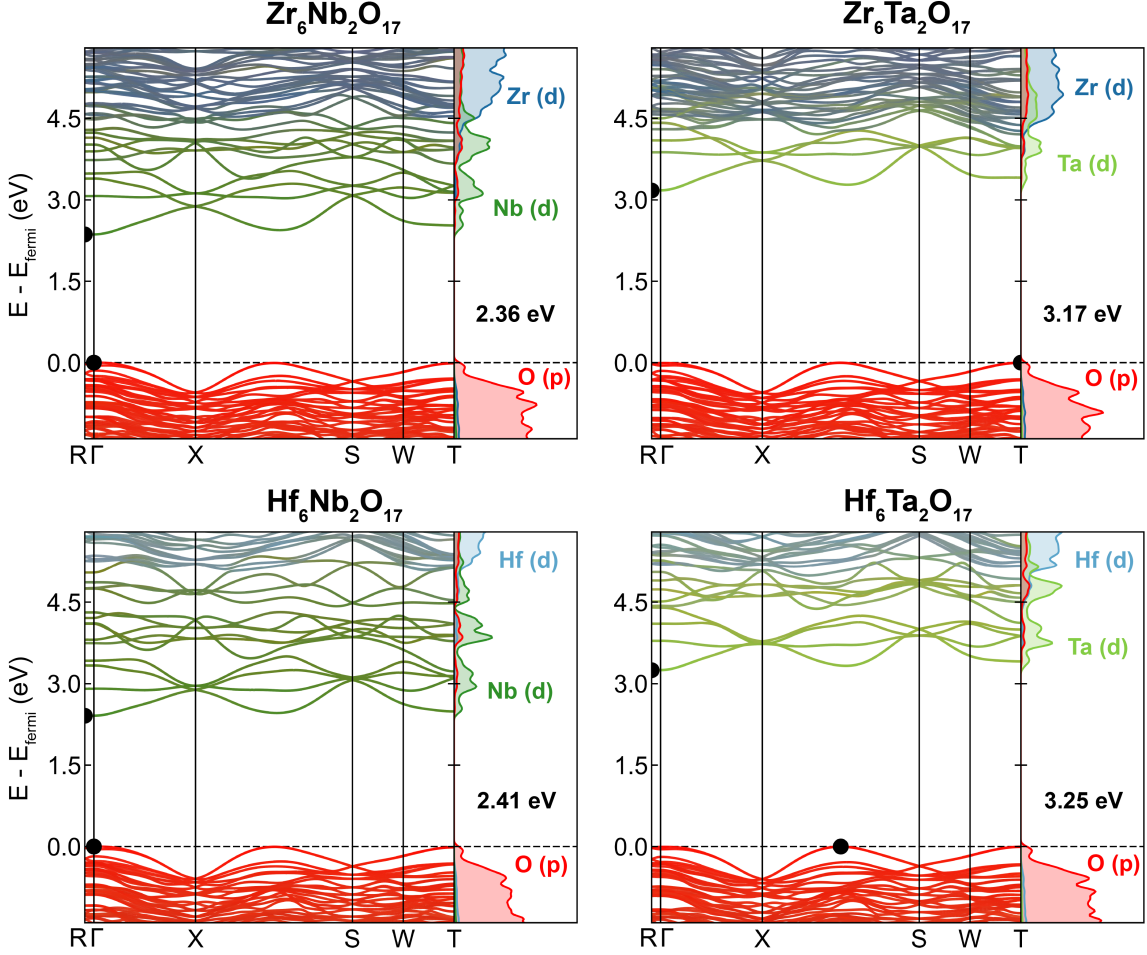

Figure 3. Projected band structures for ordered cation decorated $A_6B_2O_{17}$ ($A$ = Zr/Hf, $B$ = Nb/Ta) unit cells. Band edges are denoted with black circles for valence band maxima and conduction band minima. Only the $A$- and $B$-cations d-orbitals and oxygen p-orbitals are shown in corresponding density of states for clarity. A Gaussian smoothing (0.05 eV) is applied to the density of states, and electronic band gaps are indicated in black text.

We select the decomposition enthalpy ($\Delta H_{decomp}$) [32] to evaluate $A_6B_2O_{17}$ thermodynamic stability as this quantity leverages open-source DFT databases such as the Materials Project [33] to evaluate the lowest energy competing reaction within the relevant chemical space:

$$\Delta H_{decomp} = E_{rxn} = E_{ABC} - E_{A-B-C}. \qquad (1)$$

$\Delta H_{decomp}$ therefore reflects stability with respect to phase separation [6,34,35]. If $\Delta H_{decomp}$ is > 0 for an arbitrary compound ABC compared to its A-B-C competing phases, the reaction can be deemed endothermic: the A-B-C competing phases are more enthalpically stable than the ABC compound. Table 1 contains computed $\Delta H_{decomp}$ and associated decomposition reactions for all four $A_6B_2O_{17}$ compositions with ordered cation decorations. $Hf_6B_2O_{17}$ compositions possess larger enthalpic penalties than their $Zr_6B_2O_{17}$ counterparts. Importantly, all four DFT-calculated $\Delta H_{decomp}$ values are positive: single-phase formation is therefore enthalpically unstable with respect to $AO_2$ and $B_2O_5$ compounds, aligning with calorimetry measurements performed by Voskanyan et al. [15]. Our calculations confirm that $A_6B_2O_{17}$ ($A$ = Zr/Hf and $B$ = Nb/Ta) formulations can be classified as entropy-stabilized materials – a positive enthalpic barrier exists that must be overcome by the $-T\Delta S$ contribution to Gibbs free energy for single-phase formation. We hypothesize that vibrational entropy differences between $A_6B_2O_{17}$ and its competing oxides ($AO_2$ and $B_2O_5$) are relatively minor as the cation-oxygen coordinations are quite similar, and we attribute the configurational entropy induced by cation disorder as the primary entropic contribution which stabilizes $A_6B_2O_{17}$ as a single phase.

Table 1. Decomposition reactions and corresponding decomposition enthalpies computed using the Materials Project [33] database: ZrO$_2$ (mp-2858-r2scan), HfO$_2$ (mp-352-r2scan), Nb$_2$O$_5$ (mp-604-r2scan), and Ta$_2$O$_5$ (mp-10390-r2scan). Experimental measurements from Voskanyan et al. are included for comparison [15].

| Decomposition Reaction | Decomposition Enthalpy ($\Delta H_{decomp}$) | | |
|---|---|---|---|
| | DFT (meV/atom) | DFT (kJ/mol) | Expt. (kJ/mol) [15] |
| Zr$_6$Nb$_2$O$_{17}$ → 0.72(ZrO$_2$) + 0.28(Nb$_2$O$_5$) | +4.39 | +10.59 | +35.52 ± 6.45 |
| Zr$_6$Ta$_2$O$_{17}$ → 0.72(ZrO$_2$) + 0.28(Ta$_2$O$_5$) | +9.76 | +23.54 | +33.64 ± 5.42 |
| Hf$_6$Nb$_2$O$_{17}$ → 0.72(HfO$_2$) + 0.28(Nb$_2$O$_5$) | +11.26 | +27.16 | +38.44 ± 6.75 |
| Hf$_6$Ta$_2$O$_{17}$ → 0.72(HfO$_2$) + 0.28(Ta$_2$O$_5$) | +16.16 | +38.99 | +42.94 ± 7.03 |

## Defect Chemistry

To explore the defect chemistry of $A_6B_2O_{17}$ materials, we begin by investigating the end-member oxide compositions that compete with $A_6B_2O_{17}$ formation: ZrO$_2$, HfO$_2$, Nb$_2$O$_5$, and Ta$_2$O$_5$. Literature indicates that oxygen vacancies are the predominant defect in all $AO_2$ and $B_2O_5$ compositions considered [31,36–38]. It has also been suggested that interstitial cations are possible in these materials [39], however we limit our computational search to vacancy point defects. We use the ground-state Materials Project r$^2$SCAN structures found in our decomposition reactions, constructing $AO_2$ 2x2x2 and $B_2O_5$ 1x2x2 supercells to minimize defect interactions with periodic images. Neutral defect formation energies ($\Delta E_{defect}$) are computed as:

$$\Delta E_{defect} = (E_{defect} + \mu_i) - \Delta E_{pristine} \qquad (2)$$

where $E_{defect}$ is the energy of the defective cell, $E_{pristine}$ is the energy of the pristine cell, and $\mu_i$ is the energy of removed element $i$ in its ground-state. For neutral point defects, global electroneutrality is maintained in the calculations, which allows us to explore how charge is compensated once defects are introduced in these systems. $\Delta E_{defect}$ for $A_6B_2O_{17}$ competing end-members are shown in Figure 4a. Cation defect formation energies (~18 eV) are more than twice as large as those for oxygen (~6 eV), confirming that oxygen vacancies are indeed the predominant point defect for competing $AO_2$ and $B_2O_5$ compositions.

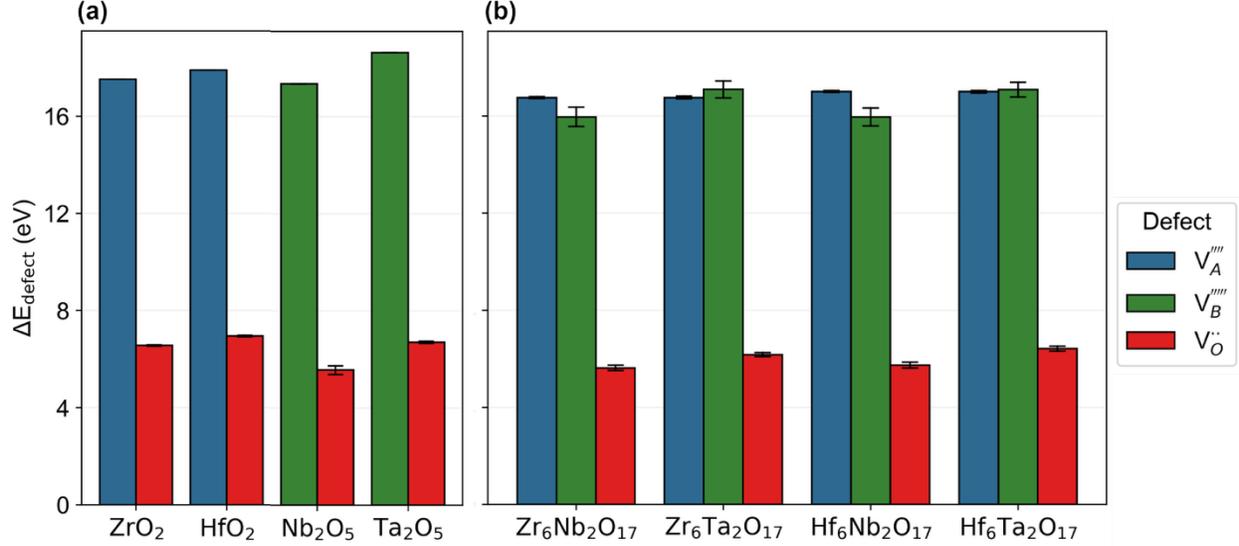

Figure 4. Calculated defect formation energies for (a) end-member oxide compositions (ZrO$_2$, HfO$_2$, Nb$_2$O$_5$, and Ta$_2$O$_5$) and (b) $A_6B_2O_{17}$ ($A$ = Zr/Hf, $B$ = Nb/Ta) ordered unit cells. Error bars indicate a 95% confidence interval.

Using the understanding afforded by the $A_6B_2O_{17}$ end-members, we expect that the most favorable point defect in $A_6B_2O_{17}$ will also be oxygen vacancies. Miruszewski et al. [23] came to a similar conclusion, suggesting that $A_6B_2O_{17}$ oxygen vacancies are compensated by Nb/Ta reduction from 5+ to 4+ in the following reaction:

$$A_6B_2O_{17} \leftrightarrow 6A_A^x + 2B'_B + V_O^{\cdot\cdot} + 16O_O^x + \frac{1}{2}O_2 \qquad (3)$$

where $O_O^X$ is the lattice oxygen ion, $V_O^{\cdot\cdot}$ is the doubly ionized oxygen vacancy, and $2B'_B$ is the reduced B ion. We continue utilizing the relaxed, ordered $A_6B_2O_{17}$ unit cells for our defect formation energy calculations to eliminate the complexity induced by cation disorder as well as to facilitate the large number of calculations required – each of the 100 atoms in the unit cell (24 $A$-cations, 8 $B$-cations, and 68 oxygen) are sequentially removed from the relaxed $A_6B_2O_{17}$ structures, resulting in 400 total defect calculations. $A_6B_2O_{17}$ $\Delta E_{defect}$ values are shown in Figure 4b. Values are quite similar to respective $AO_2$ and $B_2O_5$ end-members, and oxygen vacancies are indeed the predominant vacancy point defect within $A_6B_2O_{17}$ ($A$ = Zr/Hf, $B$ = Nb/Ta) compositions as well. We additionally explore how oxygen $\Delta E_{defect}$ values vary along the $A_6B_2O_{17}$ unit cell (Figure S3) – the most favorable oxygen $\Delta E_{defect}$ are observed near $B$-cations (occupying both 6- and 8-coordinated sites). We note, however, that while oxygen vacancies are the most likely point defect explored here, $\Delta E_{defect}$ values are quite unfavorable and suggest that oxygen point defects are unlikely to form under equilibrium synthesis conditions. To form such defects, non-equilibrium synthesis techniques may be required [21].

Having established the predominant point defect in $A_6B_2O_{17}$ as oxygen vacancies, we assess the charge compensation mechanism for the lowest energy oxygen vacancy defects using Bader charges [40]. These quantities are sensitive to changes in the electronic structure and therefore enable us to identify charge variations due to oxygen removal from the lattice. Bader charge changes ($\Delta_{Bader}$) for the most-stable oxygen defect in Zr$_6$Nb$_2$O$_{17}$, Zr$_6$Ta$_2$O$_{17}$, Hf$_6$Nb$_2$O$_{17}$, Hf$_6$Ta$_2$O$_{17}$ are shown in Figure 5. A significant change is observed for the 5+ $B$-cations (Nb/Ta) compared to the 4+ $A$-cations (Zr/Hf) in all compositions, indicating increased electron density for these cations (i.e., a reduction from 5+ to 4+ valence). Two $B$-cations gain electron density for each oxygen vacancy, confirming the charge compensation mechanism proposed in Equation 3. There is also a slight reduction observed for Zr cations in $Zr_6Ta_2O_{17}$, however, the comparatively large shift for Ta cations leads us to conclude that Ta ions primarily compensate the charge left by oxygen ion removal. The projected density of states (Figure 3) also supports this claim as $B$-cations are the lowest unoccupied states and therefore would be the most likely to accept electrons in $A_6B_2O_{17}$ compositions; note that $Zr_6Ta_2O_{17}$ also has the closest $A$-cation orbitals to the conduction band minima. While cation disorder will almost certainly introduce additional complexity into $A_6B_2O_{17}$ defect chemistry, we can conclude from our defect analysis using cation-ordered unit cells that oxygen vacancies are the most favorable point defect within $A_6B_2O_{17}$ and that their removal is charge compensated through the reduction of $B$-cations (Nb/Ta).

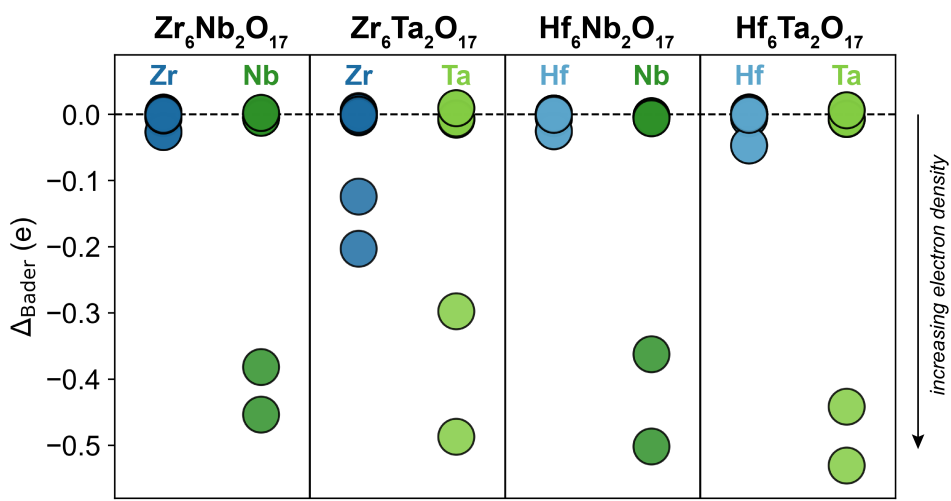

Figure 5. Changes in $A$- (Zr/Hf) and $B$-cation (Nb/Ta) Bader charges for most-stable oxygen defect formation energy in $A_6B_2O_{17}$ compositions. A negative change in Bader charge indicates a gain in electron density (i.e., a reduction in valence).

## Cation Disorder

Having explored $A_6B_2O_{17}$ compositions using the ordered cation decoration, we now investigate how cation disorder ($A$- and $B$-cations randomly occupying all 6-, 7-, and 8-coordinated sites) influences the resulting atomic and electronic structure as well as thermodynamic stability. We utilize special quasirandom structures (SQSs) for this task as they represent the high-temperature, fully-disordered limit [41] – large 1x2x2 supercells containing 400 atoms are used to explore each $A_6B_2O_{17}$ composition. We extract relaxed bond lengths from each $A_6B_2O_{17}$ structure and compare our predictions to experimentally measured bond lengths from Rietveld refinement [12]. By applying linear fits, we quantitatively compare computed bond lengths (both from the ordered unit cells and disordered SQSs) to experimental measurements: a higher quality fit should more closely align with the cation decoration present in the actual material. Disordered cation decorated SQSs achieve a highly linear fit with respect to experimental data with an $r^2 = 0.91$ compared to ordered unit cells with a lower $r^2 = 0.62$ (Figure 6). We additionally note that cation disorder induces significant displacements within the oxygen sublattice that are not observed for ordered $A_6B_2O_{17}$ unit cells (Figure S5). Similar displacements have been demonstrated for the $Mg_{1/5}Co_{1/5}Ni_{1/5}Cu_{1/5}Zn_{1/5}O$ entropy-stabilized rock salt oxide [42]. The striking agreement between our disordered SQS with experimental measurements indicates that $A_6B_2O_{17}$ ($A$ = Zr/Hf, $B$ = Nb/Ta) oxides likely have a highly disordered cation sublattice in which both $A$- and $B$-cations randomly decorate all 6-, 7-, and 8-coordinated sites.

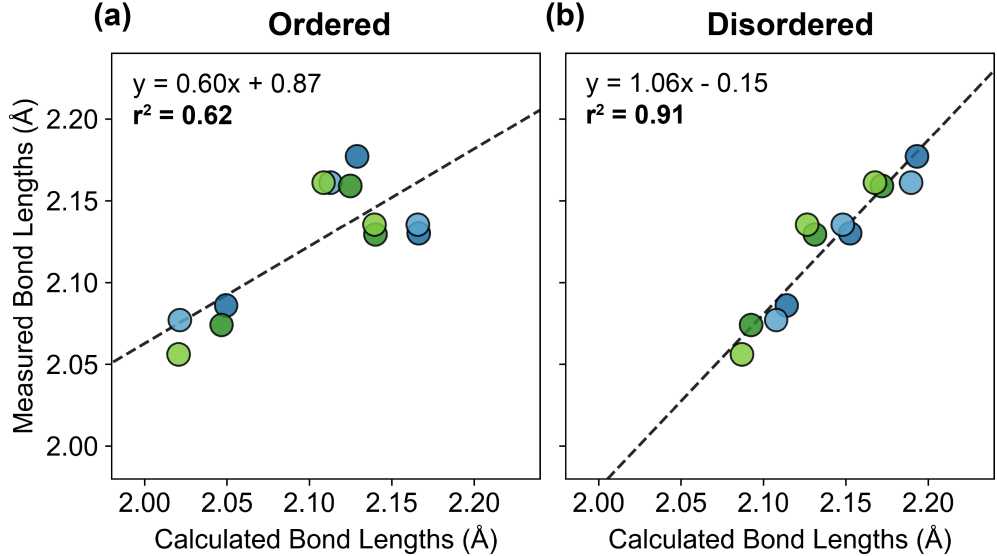

Figure 6. Bond length comparison for (a) ordered unit cell and (b) disordered SQS cation decorations with experimental measurements [12] for each cation coordination. Different $A$- and $B$-cation combinations are indicated using colors from Figure 2; a linear fit is shown as a black dashed line.

The effect of cation disorder on the $A_6B_2O_{17}$ electronic structure is shown in Figure S6. Density of states are qualitatively similar between both cation decorations, with oxygen p-orbitals remaining at the top of the valence band and $B$-cation d-orbitals at the bottom of the conduction band. Cation disorder increases the electronic band gaps by 0.37 and 0.20 eV for $Zr_6B_2O_{17}$ and $Hf_6B_2O_{17}$ compositions, respectively. Using the disordered SQS, our first-principles calculation predicts an electronic band gap of 2.73 eV for $Zr_6Nb_2O_{17}$, in excellent agreement with the experimentally measured gap of 2.78 eV from Bai et al. [31]. We attribute the increased electronic band gaps that better align with experimental measurements to displacements of the oxygen sublattice induced by cation disorder as shown in Figure S5 [43].

Decomposition enthalpies ($\Delta H_{decomp}$) for disordered $A_6B_2O_{17}$ SQSs are shown in Figure 7 as well as the ordered $A_6B_2O_{17}$ unit cells and experimental single-phase formation temperatures taken from high-temperature XRD experiments [14]. Note that decomposition reactions are identical between the ordered and disordered $A_6B_2O_{17}$ as they have the same stoichiometry (Table 1). Cation disorder induces a larger enthalpic penalty for all compositions compared to ordered unit cells by ~5.5 kJ/mol, highlighting the dynamic relationship enthalpy and entropy play within chemically disordered materials: cation disorder prevails within $A_6B_2O_{17}$ oxides despite the increased enthalpic penalty to single-phase formation. The larger $\Delta H_{decomp}$ costs associated with disordered cations also more closely align with calorimetry measurements from Voskanyan et al. [15], where first-principles $\Delta H_{decomp}$ values for disordered $Zr_6Ta_2O_{17}$, $Hf_6Nb_2O_{17}$, and $Hf_6Nb_2O_{17}$ lie within experimental error (Figure S7). The least favorable agreement compared to experiment is found for $Zr_6Nb_2O_{17}$, suggesting some discrepancy between our calculations and the calorimetry measurements. Both computed and experimental $\Delta H_{decomp}$ values are all still positive, however, demonstrating the enthalpic instability of $Zr_6Nb_2O_{17}$ with respect to $ZrO_2$ and $Nb_2O_5$. We observe a strong correlation between calculated decomposition enthalpies with the single-phase formation temperatures as noted previously for rocksalt high-entropy oxides [6], instilling confidence that DFT predictions are capable of accurately capturing relative enthalpic costs among $A_6B_2O_{17}$ compositions. Collectively, our first-principles calculations using disordered SQSs indicate cation disorder across all 6-, 7-, and 8-coordinated sites, leading to increased electronic band gaps as well as larger enthalpic barriers for single-phase formation compared to their ordered unit cell counterparts.

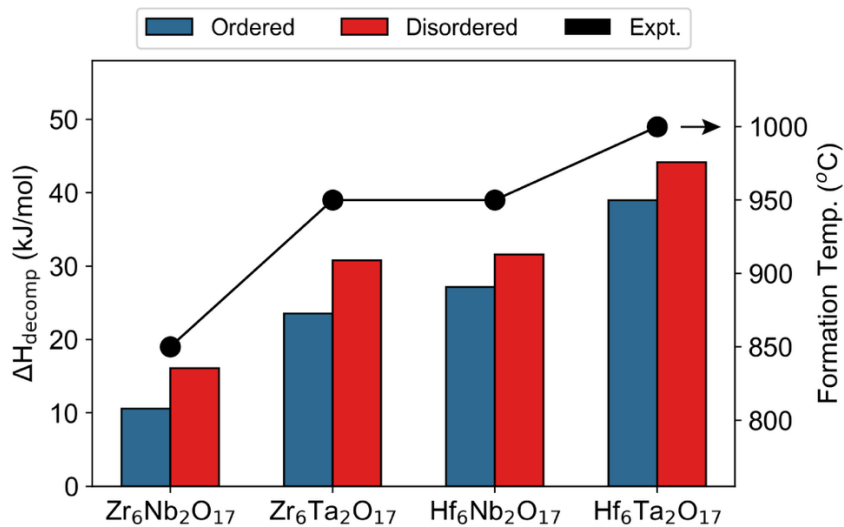

Figure 7. Decomposition enthalpy ($\Delta H_{decomp}$) comparison between ordered unit cells, disordered SQSs, and single-phase formation temperatures from experiment (via 811 peak emergence) [14] for $A_6B_2O_{17}$ ($A$ = Zr/Hf, $B$ = Nb/Ta) compositions.

## Conclusions

We explore cation selection and disorder within the $A_6B_2O_{17}$ ($A$ = Zr/Hf, $B$ = Nb/Ta) oxide family using first-principles calculations. While all compositions are structurally similar to one another with a narrow window of bond lengths, electronic structures are strongly dependent on $B$-cation species: $A_6Ta_2O_{17}$ compositions have ~30% larger band gap than their $A_6Nb_2O_{17}$ counterparts. Calculated decomposition enthalpies align closely with previous experimental measurements: all $A_6B_2O_{17}$ oxides are enthalpically unstable with respect to their $AO_2$ and $B_2O_5$ competing oxides and therefore are entropy-stabilized despite only containing two cations. Oxygen vacancies are the most favorable vacancy point defect where charge is compensated by the reduction of $B$-cations from 5+ to 4+ in all $A_6B_2O_{17}$ formulations. Predictions using disordered supercells more closely align with experimental measurements for local atomic structure, electronic structure, and thermodynamic stability, suggesting $A_6B_2O_{17}$ formulations contain a highly disordered cation decoration across all sites. Our findings establish first-principles DFT calculations are capable of quantitatively resolving chemical, structural, and thermodynamic differences within $A_6B_2O_{17}$ materials, and can be leveraged to accelerate the understanding, discovery, and implementation of these and other entropy-driven materials.

**Acknowledgements**
The authors acknowledge the use of facilities and instrumentation supported by NSF through the Pennsylvania State University Materials Research Science and Engineering Center [DMR-2011839]. Calculations utilized resources from the Roar Collab cluster of the Penn State Institute for Computational and Data Sciences.

**Methods**
The Vienna Ab-initio Software Package (VASP) 6.4.1 was used for density functional theory calculations with the projector augmented wave pseudopotentials v54 [44]. The regularized-restored strongly constrained and appropriately normed (r$^2$SCAN) functional was used for its improved accuracy and elimination of Hubbard U values to describe transition metal oxide systems [24,45]. Calculation parameters were largely unchanged from defaults of the Materials Project *MPScanRelaxSet* for r$^2$SCAN calculations, enabling our calculations to be compatible with their extensively populated materials database [33,46]. Relaxations for ordered cation unit cells allowed the cell shape, volume, and atom positions to relax (ISIF = 3), while for defect calculations the relaxed cell geometry was used and only atom positions were relaxed (ISIF = 2). Forces were minimized to less than 50 meV/Å for defect calculations to facilitate the large number of calculations, and 20 meV/Å otherwise. Disordered cation decorated structures were constructed as 400-atom (1x2x2 supercell) special quasirandom structures (SQSs) [41]. The relaxed cell shape of the ordered unit cells are used as inputs, however both the atom positions and cell volume were allowed to relax (ISIF = 8). A KSPACING value of 0.25 was used to ensure convergence in all cases. Non-spin polarized calculations (ISPIN = 1) were used for all calculations other than defect calculations (where unpaired electrons may be present); defect calculations (with ISPIN = 2) were initialized as ferromagnetic. Calculations were managed and analyzed using the Custodian, Pymatgen [47], and Sumo [48] packages. SQSs were generated using the Integrated Cluster Expansion Toolkit (ICET) [41,49] and Bader charge analysis was performed using code from the Henkelman group [40].

# Supplemental Material

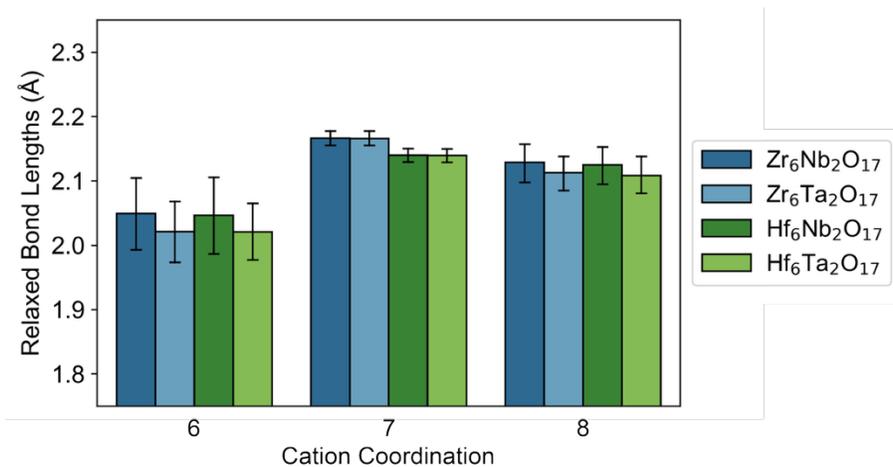

Figure S1. Relaxed unit cell bond length predictions for ordered $A_6B_2O_{17}$ ($A$ = Zr/Hf, B = Nb/Ta) unit cells across all 6-, 7-, and 8-coordinated cation sites. Errors bars indicate a 95% confidence interval.

Table S1. Average Bader charges for ordered $A_6B_2O_{17}$ 100-atom unit cells and competing phases identified in our decomposition reactions: $ZrO_2$, $HfO_2$, $Nb_2O_5$, and $Ta_2O_5$. Larger values indicate more ionic-like bonding.

| Composition ($A_6B_2O_{17}$) | Bader Charge | | |
|---|---|---|---|
| | $A$-Cation | $B$-Cation | Oxygen |
| $Zr_6Nb_2O_{17}$ | +2.69 | +2.80 | -1.28 |
| $Zr_6Ta_2O_{17}$ | +2.69 | +2.96 | -1.30 |
| $Hf_6Nb_2O_{17}$ | +2.67 | +2.80 | -1.27 |
| $Hf_6Ta_2O_{17}$ | +2.67 | +2.96 | -1.29 |
| $ZrO_2$ | +2.68 | -- | -1.34 |
| $HfO_2$ | +2.65 | -- | -1.33 |
| $Nb_2O_5$ | -- | +2.80 | -1.12 |
| $Ta_2O_5$ | -- | +2.96 | -1.20 |

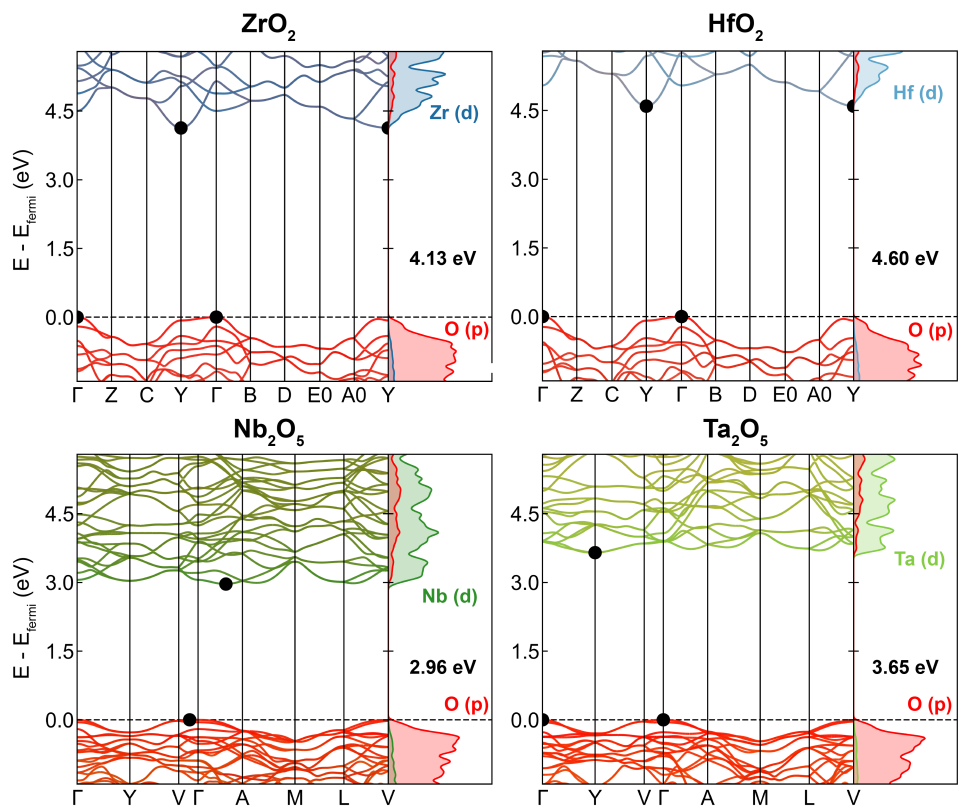

Figure S2. Projected band structures for competing phases identified in our decomposition reactions: $ZrO_2$, $HfO_2$, $Nb_2O_5$, and $Ta_2O_5$. Band edges are denoted with black circles for valence band maxima and conduction band minima. Only the *A*- and *B*-cations d-orbitals and oxygen p-orbitals are shown in corresponding density of states for clarity.

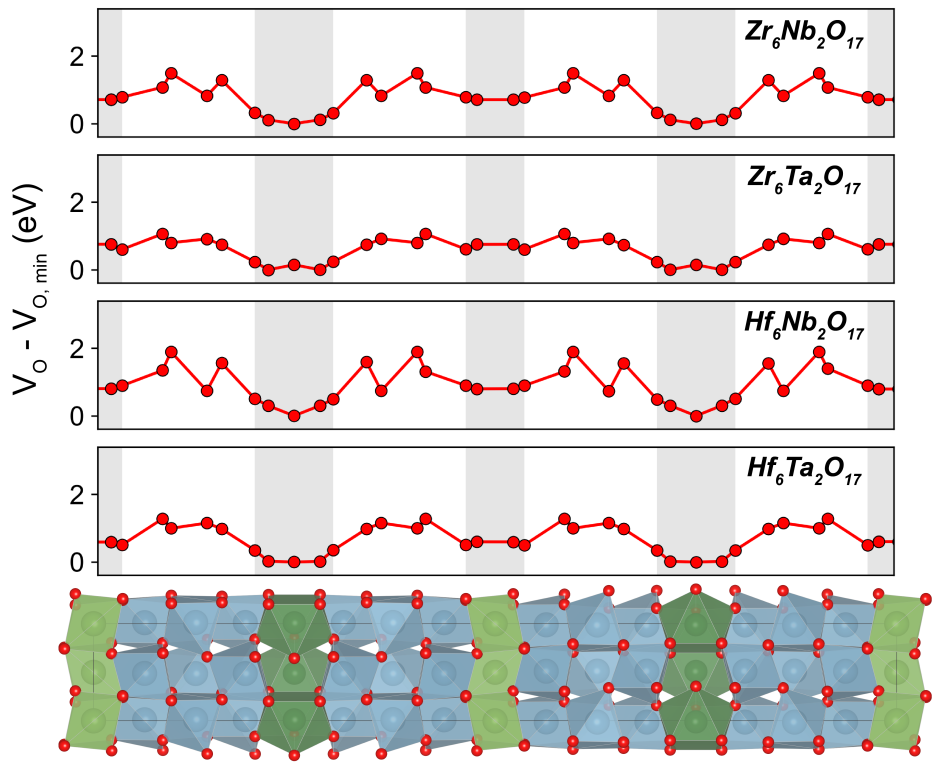

Figure S3. Relative oxygen defect formation energy as a function of position in ordered $A_6B_2O_{17}$ unit cell. Grey shaded areas indicate 6- and 8-coordinated sites on which $B$-cations (Nb/Ta) reside. Figure 1 is reproduced below for clarity.

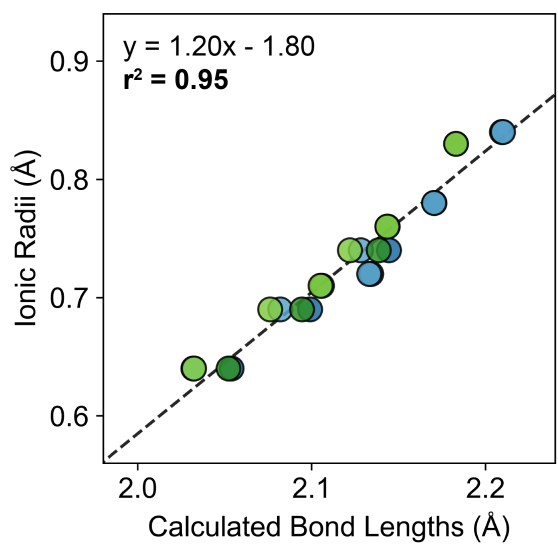

Figure S4. Bond length comparison for disordered SQS cation decorations with Shannon-Prewitt ionic radii [28] for each cation coordination. Different $A$- and $B$-cation combinations are indicated using colors from Figure 2; a linear fit is shown as a black dashed line.

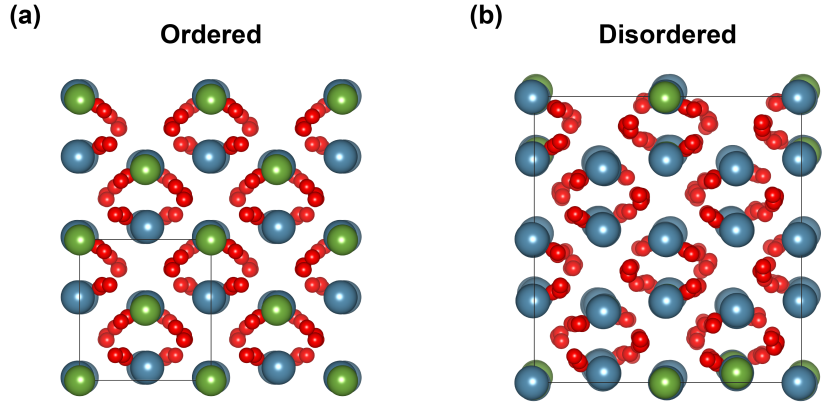

Figure S5. Relaxed atomic structures for (a) ordered unit cell and (b) disordered SQS supercell structures looking down [010] direction. *A*-cations are shown in green, *B*-cations in blue, and oxygen anions in red. Notice the significant distortion cation disorder induces on the oxygen sublattice.

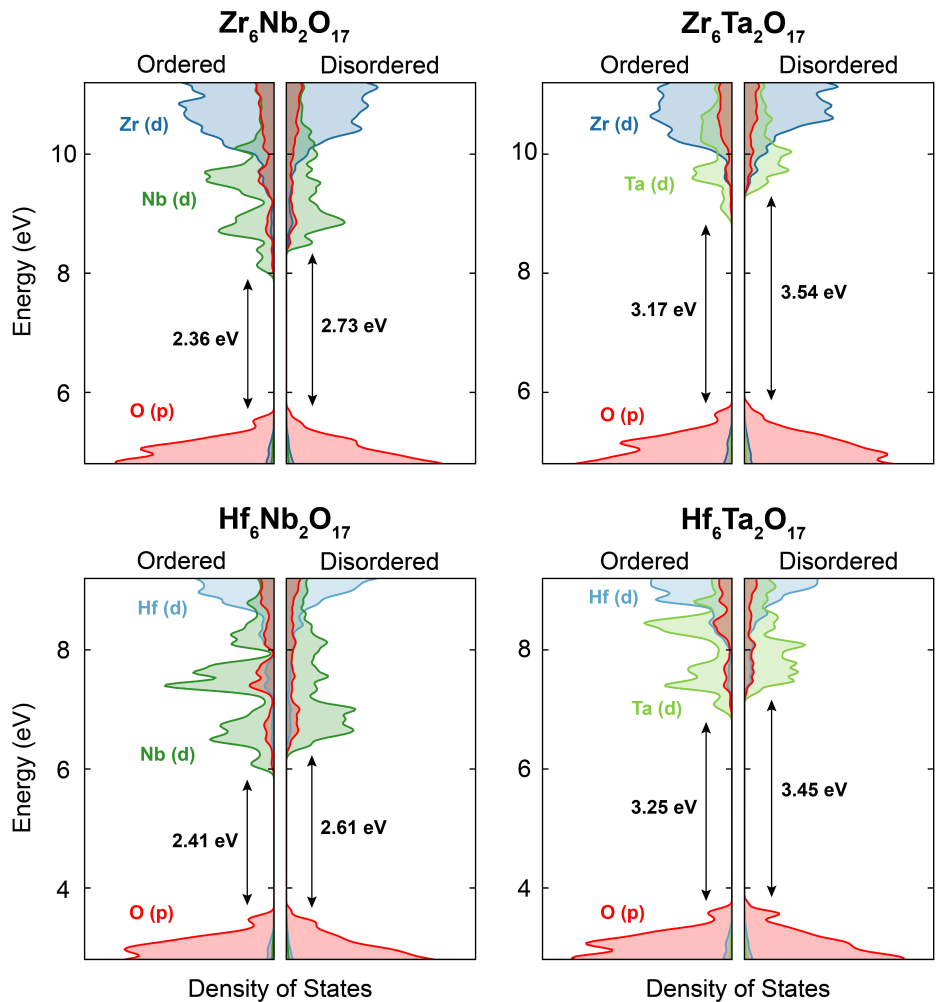

Figure S6. Density of states for $A_6B_2O_{17}$ (*A* = Zr/Hf, *B* = Nb/Ta) ordered unit cells and disordered SQSs. Only the *A*- and *B*-cations d-orbitals and oxygen p-orbitals are shown for clarity. A Gaussian smoothing (0.05 eV) is applied to the density of states and electronic band gaps are indicated in black.

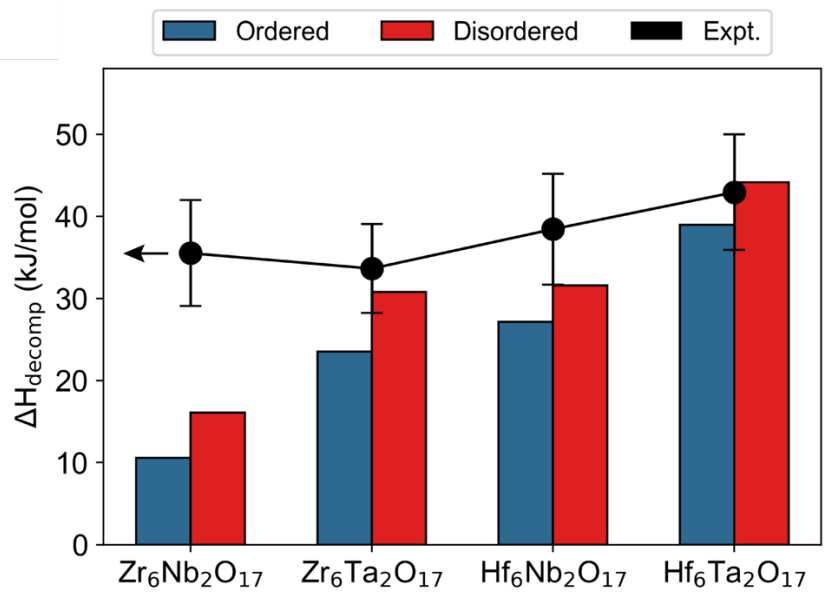

Figure S7. Decomposition enthalpy ($\Delta H_{decomp}$) comparison between ordered unit cells, disordered SQSs, and experimental calorimetry measurements [15] for $A_6B_2O_{17}$ ($A$ = Zr/Hf, $B$ = Nb/Ta) compositions.